\documentclass[a4paper,twocolumn,11pt,unpublished]{quantumarticle}
\pdfoutput=1

\usepackage[utf8]{inputenc}
\usepackage[english]{babel}
\usepackage[T1]{fontenc}
\usepackage{hyperref}
\usepackage{physics}
\usepackage{tikz}
\usepackage[normalem]{ulem}
\usepackage{latexsym,graphicx,amsmath,amsfonts,amssymb,url,textcase,bm}

\DeclareMathOperator*{\argmax}{argmax}
\newcommand{\km}[1]{{\color{black}{#1}}}

\begin{document}
\title{Benchmarking the algorithmic performance of near-term neutral atom processors}
\author{K. McInroy}
\email{kathryn.mcinroy.2018@uni.strath.ac.uk}
\affiliation{Department of Physics and SUPA,University of Strathclyde, Glasgow G4 0NG, United Kingdom}
\author{N. Pearson}
\orcid{0000-0002-5993-7258}
\affiliation{Department of Physics and SUPA,University of Strathclyde, Glasgow G4 0NG, United Kingdom}
\author{J.D. Pritchard}
\orcid{0000-0003-2172-7340}
\affiliation{Department of Physics and SUPA,University of Strathclyde, Glasgow G4 0NG, United Kingdom}

\begin{abstract}
Neutral atom quantum processors provide a viable route to scalable quantum computing, with recent demonstrations of high-fidelity and parallel gate operations and initial implementation of quantum algorithms using both physical and logical qubit encodings. In this work we present a characterization of the algorithmic performance of near term Rydberg atom quantum computers through device simulation to enable comparison against competing architectures. We consider three different quantum algorithm related tests, exploiting the ability to dynamically update qubit connectivity and multi-qubit gates. We calculate a quantum volume of $\mathbf{\mathit{V_{Q}}=2^{9}}$ for 9 qubit devices with realistic parameters, which is the maximum achievable value for this device size and establishes a lower bound for larger systems. We also simulate highly efficient implementations of both the Bernstein-Vazirani algorithm with >0.95 success probability for 9 data qubits and 1 ancilla qubit without loss correction, and Grover's search algorithm with a loss-corrected success probability of 0.97 for an implementation of the algorithm using 6 data qubits and 3 ancilla qubits using native multi-qubit $\mathbf{CCZ}$ gates. Our results indicate  Rydberg atom processors are a highly competitive near-term platform which, bolstered by the potential for further scalability, can pave the way toward useful quantum computation.
\end{abstract}

\maketitle

\section{Introduction} 
Quantum computing has the potential to offer a wide range of applications which, in the limit of a universal fault-tolerant architecture, can provide speed-up of classically hard algorithms such as factorization \cite{shor97}, optimization \cite{zhu22,wurtz22} and machine learning \cite{biamonte17}, and enable efficient calculation of properties of quantum systems relevant for problems in quantum chemistry and material science \cite{cao19,bauer20,mcardle20}. These promising applications have driven significant {\color{black}development} in both industry and academia towards scalable architectures for quantum computing based on superconducting circuits \cite{kjaergaard20}, trapped ions \cite{georgescu20,srinivas21,bruzewicz19,pino21} and photonic circuits \cite{slussarenko19}. However, current systems typically operate in the noisy intermediate-scale quantum (NISQ) regime \cite{preskill18} limiting their near-term utility.

To quantify the performance of {\color{black} digital} quantum hardware and facilitate cross-platform comparison, a range of gate and circuit level benchmarks have been developed. Techniques for evaluating gate-level performance include randomized benchmarking \cite{knill08,cross16}, quantum process tomography \cite{merkel13, giamoco23} and gate set tomography \cite{blumekohout17}, which provide valuable tools for characterising the performance of few-qubit operations, but {\color{black} cannot be used deterministically to quantify} the ability of a given device to perform a useful quantum algorithm \cite{murphy19}. Instead, algorithmic \cite{michielsen17,boxio18,wright19} and volumetric benchmarking \cite{cross19,blumekohout20} metrics have been developed to enable characterization of the device in regimes relevant for implementing a quantum algorithm. Early work focused on side-by-side comparison for the performance of different devices, such as comparing 5 qubit trapped ion and superconducting circuit devices \cite{linke17}, however more sophisticated single-value metrics such as quantum volume \cite{cross19} or application benchmarking \cite{mills21,lubinski23} provide a {\color{black} metric for} the potential utility of a given quantum processor. Already, quantum volume has been used to characterize both superconducting circuits and trapped ion devices, with peak values of $V_{Q}=2^{9}$ \cite{Gambetta22TWT} and $V_{Q}=2^{16}$ \cite{moses23} respectively.

 Recently, neutral atom quantum processors based on arrays of atoms trapped in individual optical tweezers, as illustrated in Fig.~1, have emerged as a promising technology for developing scalable quantum computing \cite{morgado21,adams20,henriet20}. This architecture supports arrays of {\color{black}over} $1000$ identical and high-quality qubits \cite{huft22}, capable of implementing high fidelity single-qubit \cite{sheng18,nikolov23}, two-qubit \cite{levine19} and multi-qubit gate operations \cite{levine19,evered23,pelegri22} in parallel {\color{black} by} exploiting the strong long-range interactions between highly excited Rydberg states to provide controllable couplings between qubits \cite{saffman10}, with state-of-the-art two-qubit gate fidelities reaching 99.5\% \cite{evered23,ma23}. Small-scale {\color{black} digital} quantum processors {\color{black} based on Rydberg atoms} have been used to demonstrate quantum algorithms on up to 5 qubits \cite{graham22} and generate topological encodings by exploiting coherent atomic transport to  engineer non-local qubit connectivity \cite{bluvstein22}. The ability to dynamically rearrange qubits provides a new modality for efficiently implementing quantum algorithms beyond the typical planar architectures \cite{tan23}, and provides a route to future fault-tolerant operation using quantum error correction \cite{auger17,cong22,sahay23}. Recent work demonstrating transverse gates on 48 logically encoded qubits highlights the viability and  capability of the neutral atom approach \cite{bluvstein23}. This architecture also permits analogue quantum computation for simulation of quantum systems \cite{scholl21,ebadi21,semeghini21}, implementation of graph-based optimization problems \cite{pichler18,pichler18a,nguyen23,cain23,ebadi22}, and integer factorization \cite{park23}.

\begin{figure}[t!]
         \includegraphics[width=\columnwidth]{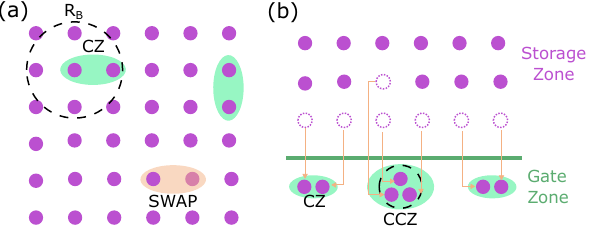}
    \caption[Array and Blockade]{\label{fig:fig1}{Schematic of different modalities of a Rydberg atom array quantum computer. Dashed black circles identify the blockade radius $R_\mathrm{B}$ of the the central qubit. (a) A static architecture with nearest-neighbour connectivity mediated by the Rydberg blockade mechanism, with gates implemented using local addressing lasers. The transparent green and orange ovals represent $CZ$ and SWAP gates respectively. (b) A reconfigurable architecture which utilizes mobile tweezers to coherently move qubits\km{, shown as orange arrows,} between a storage zone and a gate zone enabling parallel $CZ$ or $CCZ$ gates. We assume the capability of fast single-site addressing for single qubit gates in both modes.}}
\end{figure} 

In this paper we numerically characterize the potential performance of a neutral atom quantum computer, considering both volumetric and algorithmic benchmarks to facilitate comparison of current state-of-the-art hardware against competing platforms and evaluate the potential for utility in near-term applications. Specifically, we simulate performance using {\color{black}a} virtual quantum device \cite{gustiani23}, {\color{black} wrapping} the Quantum Exact Simulation Toolkit (QuEST) {~\cite{jones19}},{\color{black} and} using device parameters motivated by recent experimental demonstrations. \km{We focus on virtual devices of 9 or 10 qubits due to hardware limitations on the simulation.} We consider three benchmarks; firstly, quantum volume, which evaluates the largest device size for which gate depth matches qubit count when preparing random output states using one-qubit and two-qubit gates on the hardware. We compare {\color{black}the }performance of devices using either fixed or reconfigurable qubit modalities, and consider two different error models for implementing controlled-Z gates \cite{levine19,pelegri22}, to investigate system performance. Our simulations yield an upper bound {\color{black} on} {\color{black}the} quantum volume of $V_{Q}=2^{9}$ for a 9 qubit Rydberg atom device, which is the maximum attainable value for a device of this size and hence a lower bound on what we expect to see with an implementation of this procedure on a realised system with many more available qubits. 

One limitation of the quantum volume metric is {\color{black} that} it fails to exploit the {\color{black} hardware-specific} benefits for implementing more general algorithms, \km{such as the ability to} perform \km{native} multi-qubit gates. To address this we consider the performance of the the Bernstein-Vazirani algorithm {~\cite{bernstein97}}, showing a high success rate for a 9+1 qubit algorithm implemented via coherent qubit rearrangement. Additionally, we demonstrate the improved performance of using native multi-qubit gates in implementing Grover's search algorithm {~\cite{grover97}}, showing a successful implementation of this algorithm on a 64-item search space using a 9 qubit virtual device. These results highlight the potential for near-term Rydberg atom processors to be competitive against alternative technologies, with a key advantage in the ease of scaling to high qubit counts for future implementation of quantum error correction. 

The manuscript is arranged as follows. In Section~\ref{sec:VQD} we give an overview of the virtual neutral atom quantum device used for simulation including realistic parameters for error rates and outline the different native gate schemes and their associated error channels. In Section~\ref{sec:benchmark} we present the results of the different volumetric and algorithmic benchmarks, including discussion of assumptions around qubit connectivity. Finally, in Section~\ref{sec:discussion} we provide a summary of our {\color{black}findings}  and a future outlook.

\section{Virtual neutral atom processor}
\label{sec:VQD}

\subsection{Device Architecture}
We consider a neutral atom processor based on reconfigurable arrays of individual atoms, with qubit levels encoded onto the hyperfine ground states and gates implemented by coupling atoms to highly-excited Rydberg states \cite{saffman10,adams20}. Rydberg atoms experience strong, long-range dipole-dipole interactions which leads to a blockade effect preventing more than a single Rydberg excitation for atoms within a radius $R_\mathrm{b}\lesssim5-10~\mu$m. This blockade effect can be exploited to enable native implementation of a controlled-phase ($CZ$) gate for two qubits \cite{jacksch00}, and $CCZ$ for three qubits \cite{levine19}.

We model two different modalities for operating the neutral atom device - the first shown in Fig.~\ref{fig:fig1}(a) assumes a static spatial configuration and the ability to apply sequential and addressable gate operations between any pair of neighbouring atoms as recently demonstrated by Graham \emph{et al.} \cite{graham22} for implementing small-scale algorithms on up to 5 qubits. \km{For this static configuration, to connect non-neighbouring qubits we implement SWAP gates based on three $CZ$ operations, and minimize the number of SWAP gates required by leaving the qubits in the new configuration until an additional non-connected 2-qubit gate is required.} The second approach is based on using reconfigurable atom arrays with dynamic qubit connectivity inspired by Bluvstein \emph{et al.} \cite{bluvstein22} to allow parallel two and multi-qubit gate operations to be implemented between different qubits at each stage of the circuits by \km{moving qubits between a storage and gate zone using mobile tweezer traps}, as shown in Fig.~\ref{fig:fig1}(b).

\subsection{Two and Three Qubit Gate Protocols}
Throughout this work we consider two specific implementations of the $CZ$ gate \km{to analyse the relative impact of different gate errors on algorithmic performance}: a procedure based on two-photon adiabatic rapid passage (ARP) \cite{pelegri22} ($CZ_{\mathrm{ARP}}$) with a theoretical gate fidelity \cite{pelegri22} $\mathcal{F}$=99.81\%, and the Levine-Pichler (LP) procedure based on global excitation of qubits \cite{levine19} ($CZ_{\mathrm{LP}}$) with a theoretical gate fidelity $\mathcal{F}$=99.87\%. The operator representations of each gate are calculated by numerical simulation of the atomic energy levels, taking into account loss from spontaneous emission and the finite blockade radius using the approach described in Ref.~\cite{pelegri22} resulting in the following non-unitary two-qubit gate operators where \km{loss from the computational basis due to spontaneous decay is represented by non-unity amplitudes on the diagonal terms, and phase errors due to imperfect blockade result in deviations from the values of $0$ or $\pi$ expected for an ideal gate}.
\begin{align}
\label{eq:CZGates1}
&U_{CZ_{\mathrm{ARP}}}=\ketbra{00}+0.9990e^{0.9906i\pi}(\ketbra{01}\nonumber\\&\quad+\ketbra{10})+0.9986e^{1.000i\pi}\ketbra{11},\\
&U_{CZ_{\mathrm{LP}}}=\ketbra{00}+0.999320e^{-0.013i\pi}(\ketbra{01}\nonumber\\&\quad+\ketbra{10})
+0.999458e^{0.985i\pi}\ketbra{11}.
\label{eq:CZGates2}
\end{align}

Whilst both gate protocols achieve similar theoretical fidelities, \km{the ARP gate has a larger amplitude error due to increased probability for spontaneous decay and leakage during the gate, whilst the Levine-Pichler gate using shorter gate pulses resulting in suppressed loss but a larger error in the accumulated phase on the $\ket{11}$ term due to increased blockade leakage error.}

 \km{To capture a fuller picture of the capability of neutral atom devices with native multi qubit gates we} also consider a 3-qubit controlled-controlled-Z ($CCZ_{\mathrm{ARP}}$) phase gate realized by the two-photon adiabatic rapid passage technique \km{with a theoretical fidelity $\mathcal{F}=99.54\%$}{~\cite{pelegri22}}. The operator representation of this gate is by
\begin{align}
\label{eq:CCZGate}
&U_{CCZ_{\mathrm{ARP}}}=\ketbra{000}+0.9981e^{0.9845i\pi}(\ketbra{001}\nonumber
\\&\quad+\ketbra{010}+\ketbra{100})\nonumber+0.9973e^{0.9934i\pi}\nonumber
\\&\quad(\ketbra{011}+\ketbra{101}+\ketbra{110})\nonumber
\\&\quad+0.9963e^{0.9911i\pi}\ketbra{111}.
\end{align}

\subsection{Virtual Quantum Device}
Our neutral atom processor is modelled using the virtual quantum device library \cite{gustiani23} for (QuEST)~\cite{jones19}, which provides built in methods for modelling realistic error sources using Kraus operators, and provides a framework for encoding and rearranging qubit geometry including evaluating which gates are allowed based on qubit proximity. Single qubit errors are encoded by defining relaxation ($T_1=4$~s) and depolarization ($T_2=1.49$~s) times based on \cite{bluvstein22}, and we include a state preparation error of $\epsilon_\mathrm{Init}=0.003$ to account for finite optical pumping. Table~\ref{tab:param} contains a complete list of parameters used for modelling the virtual quantum device.

For the quantum volume and Bernstein-Vazirani algorithms we first generate the circuits using QisKit, and then transpile for the virtual device by converting to the native $CZ$ and single qubit gate sets. The resulting circuits are then imported into the QuEST library for modelling device performance.

\km{To model} the dynamically reconfigurable qubit approach of Fig.~\ref{fig:fig1}(b), we identify suitable routing approaches to provide efficient implementations of the required circuits, and model the movement as {\color{black}taking} a fixed duration, $\tau$, using the parameters from Ref.~\cite{bluvstein22} to give typical movement times required. {\color{black}For the quantum volume calculation and Grover's search algorithm, where multiple atoms may need to be moved, we use $\tau = \tau_{\mathrm{move}^\mathrm{A}}$. For the Bernstein-Vazirani algorithm implementation, where only a single atom must be moved in each reconfiguration, we use $\tau = \tau_{\mathrm{move}^\mathrm{B}}$.} This ensures realistic estimation of error expected in the regime of implementing qubit reconfiguration, without being dependent on the proposed routes being optimal.  

For each algorithm we model the evolution {\color{black} of the full system and} extract the {\color{black}final} density matrix. {\color{black}Because we model} loss as a non-unitary gate process, the resulting density matrix is not normalized, providing an estimate of the total probability of loss from the computational basis during the algorithm{\color{black}. We then} perform re-normalization of the output to extract a loss-corrected expectation. Practically, this loss-correction process can be achieved in a real experiment by implementing non-destructive readout, as demonstrated in Ref.~\cite{kwon17,nikolov23}, to determine both the qubit state and to check if an atom was lost from the tweezer, allowing post-selecting of data only from cases where all qubits \km{are retained and remain in the computational basis}.

\begin{figure*}[t!]
     \centering    
         \includegraphics[width=\textwidth,keepaspectratio]{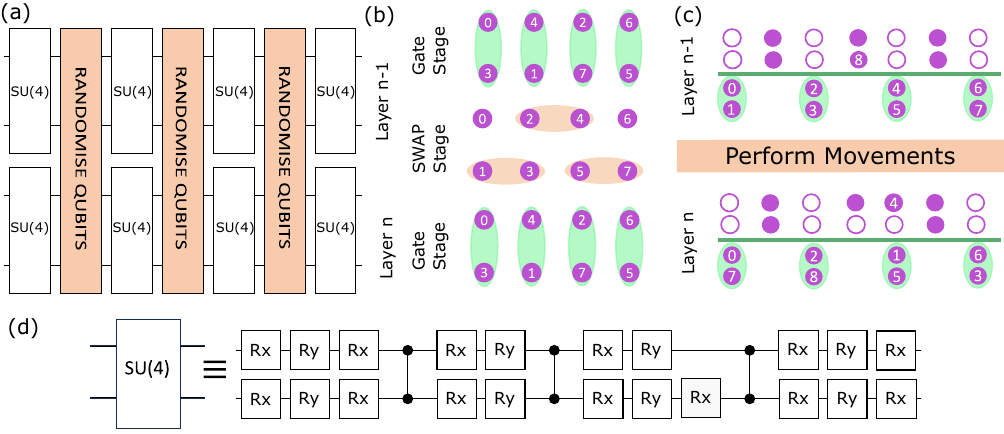}
    \caption{\label{fig:QuantumVolumeInfo}{A schematic for implementing quantum volume calculation on a Rydberg atom processor. (a) \km{We implement circuits of this form, consisting of layers of Haar uniform $SU(4)$ gates applied to pairs of qubits where between each layer we randomly assign qubit pairs.} (b) The connectivity procedure for the static mode is illustrated. \km{Qubits are initially numbered such that the first layer of $SU(4)$ gates is performed between atom pairs. The qubits are then randomly reordered by applying SWAP gates along the rows, before applying the subsequent layer of $SU(4)$ gates. This is repeated until the full circuit depth is reached.} (c) The reconfigurable connectivity procedure involves physically reordering the qubit pairs within the gate zone to allow for parallel two-qubit gate implementation. \km{The $SU(4)$ gates are implemented in parallel, meaning only one set of movements is necessary  between each circuit layer}. (d) An example {\color{black} of an }$SU(4)$ gate decomposition {\color{black}into a native gate set consisting of} 21 arbitrary rotations around the $x$ and $y$ axes of the Bloch sphere {\color{black}and 3 CZ gates}.}}
\end{figure*}

\section{Benchmarking Algorithmic Performance}
\label{sec:benchmark}
\subsection{Quantum Volume}
We first look at quantum volume as a measure of the algorithmic performance of our simulated device \cite{cross19}. This standardized form of benchmarking is a measure of the largest square circuit of arbitrary \km{two-qubit $SU(4)$ gates} that can be reliably implemented on a device{\color{black}, thus requiring that the number of qubits, $N_Q$, is equal to the circuit depth, $d$. The procedure for calculating this value is well documented in \cite{aaronson16} and involves } simulating a circuit of the form shown in Fig.~\ref{fig:QuantumVolumeInfo}(a), both on our virtual quantum device and for an ideal, error-free system. With these circuits we evaluate the heavy output problem{\color{black}. This involves} calculating the probability of the error-prone virtual device returning a ``heavy output'', defined {\color{black} as the set of} output bit strings following an ideal (without any error) implementation of the circuit, which have measurement probabilities above the median probability of generating any given bit string from the full output distribution. 

We perform this procedure on the virtual device using both $CZ_{\mathrm{ARP}}$ and $CZ_{\mathrm{LP}}$ protocols as defined in Eqs.~\ref{eq:CZGates1} and \ref{eq:CZGates2} for the two different modalities, static moves with SWAP operations and dynamically reconfigurable tweezers using coherent movement between gates. {\color{black}The quantum volume circuit consists of layers} of random $SU(4)$ unitaries, {\color{black}interleaved with operations which randomly rearrange} the qubit pairs, {\color{black} either} by applying SWAP operations along the rows to re-order odd and even qubit pairs {\color{black} in the static configuration device}, or by simply moving qubits into the relevant positions {\color{black}in the reconfigurable case. This is} illustrated in Fig.~\ref{fig:QuantumVolumeInfo}(b) and (c). For both approaches, the random $SU(4)$-gates are decomposed into three $CZ$ and up to 15 single-qubit gate rotations, as shown in Fig.~\ref{fig:QuantumVolumeInfo}(d).

For each circuit depth, we calculate the mean of the probability of obtaining a heavy output $\Bar{h}_{p}$ across 200 randomly generated circuits. Following the method of Ref.~\cite{cross19}, the criteria of success for this metric is that the $\Bar{h}_{p}>\frac{2}{3}$ by a one-sided confidence interval of twice the standard error of the mean. From these results, we can then calculate the quantum volume of the device
\begin{equation}
\label{eq:QuantumVolume}
    \log_{2}V_{Q}=\argmax_{N_{Q}} min(N_{Q},d).
\end{equation}

\begin{figure}[t!]
     \centering    
    \includegraphics[width=\columnwidth,keepaspectratio]{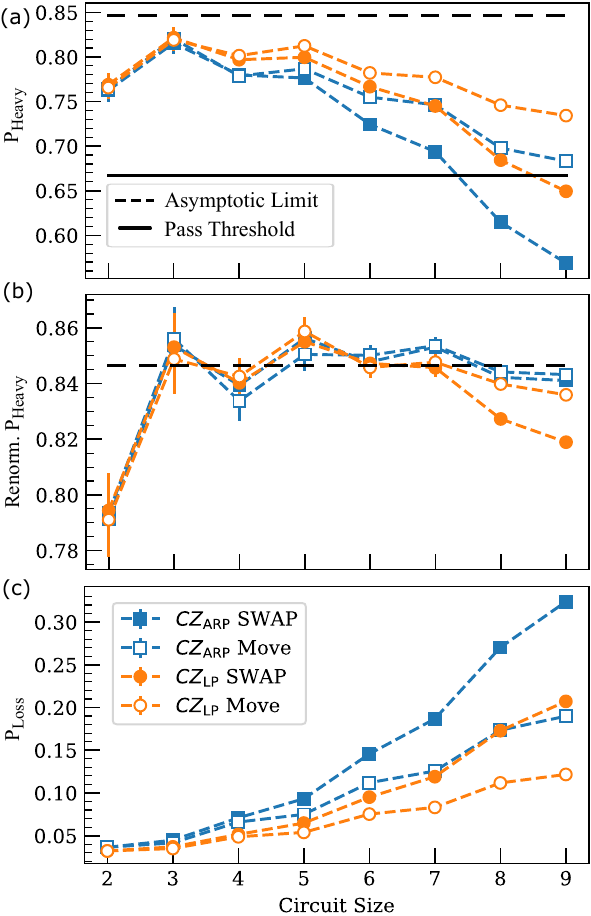}
    \caption[]{\label{fig:QV} Benchmarking Quantum Volume for square circuits up to $9\times9$ for both $CZ$ gates with both connectivity schemes for between 2$\mathrm{x}2$ and $9\mathrm{x}9$ square circuits. (a) The unprocessed average $\Bar{h}_{p}$ for each layer. (b) The corrected $\Bar{h}_{p}$ when post-selection for survival is performed. (c) The mean loss probability $P_{\mathrm{Loss}}$ at each size of circuit.}
\end{figure}

The resulting heavy output $\Bar{h}_{p}$ and loss probabilities $P_{\mathrm{Loss}}$ for each implementation are shown in Fig.~\ref{fig:QV}. These results show that, for both gate protocols, the SWAP based static modality performs worse due to the larger number of required $CZ$ gates{\color{black}. Without post-selection the setup with the lowest quantum volume is for the static modality with the ARP gate, which has more loss than the LP gate, and results in a quantum volume of $V_Q=2^7$. When using the reconfigurable modality, both gate types result in a quantum volume of $V_Q=2^9$, the maximum attainable for the device size we consider, and the circuit using LP gates has the best performance, with $P_{\mathrm{Loss}}<0.15$ for a $9\times9$ square circuit.} 

In the regime where data is post-selected for atom loss using a non-destructive imaging sequence as described above, the renormalized heavy output probabilities in Fig.~\ref{fig:QV}(b) show that actually the ARP-based gates with coherent movements provide the best corrected performance due to the lower phase errors, and that all four combinations yield a corrected $V_Q$ at $N_Q=9$ close to the \km{asymptotic limit of the heavy output probability} of \km{0.851}.

These predicted results are already comparable to recently reported numbers for superconducting qubit systems with $V_{Q}=2^{9}$ on the IBM Prague processor \cite{Gambetta22TWT}, whilst state-of-the-art across all hardware platforms comes from trapped ion systems offering $V_{Q}=2^{16}$ {~\cite{moses23}}, demonstrating that neutral atom processors offer a highly competitive approach to scalable computing.  \km{Note our simulations provide a lower bound on the attainable quantum volume of Rydberg atom devices of $V_{Q}\ge2^{9}$}, \km{even in the absence of post-selection, however the exact threshold requires calculations to be performed at larger values of $N_Q$ which is beyond the scope of this current study}.

\subsection{The Bernstein-Vazirani Algorithm}
\begin{figure}[t!]
     \centering    
         \includegraphics[width=\columnwidth,keepaspectratio]{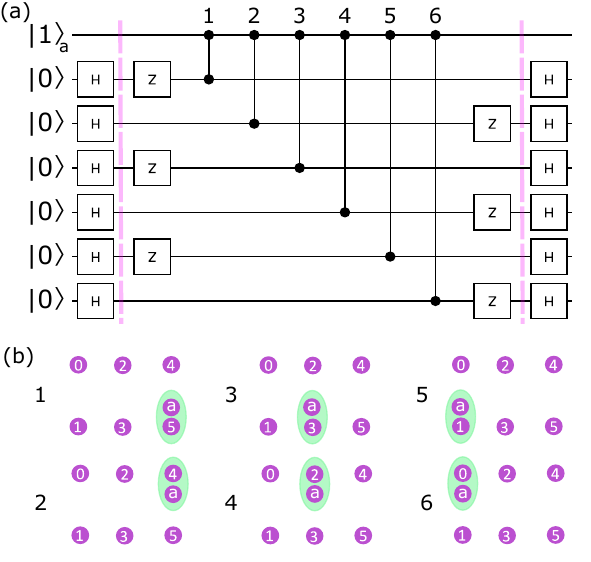}
    \caption{\label{fig:BVInfo}{A schematic for implementing the Bernstein-Vazirani algorithm on a Rydberg atom quantum computer in the reconfigurable mode. (a) The standard form of the circuit on $6+1$ qubits is shown for an oracle\km{, highlighted between the pink lines, encoding the bit string $`111111`$ and} consisting of $CZ$ gates conditioned on the data qubits targeting an ancilla qubit. (b) The movement procedure is sketched, where the ancilla qubit moves between zones where it can connect with isolated data qubits. Each $CZ$ gate requires one movement of the ancilla qubit.}}
    \end{figure}

    \begin{figure}[t!]
     \centering    
         \includegraphics[width=\columnwidth]{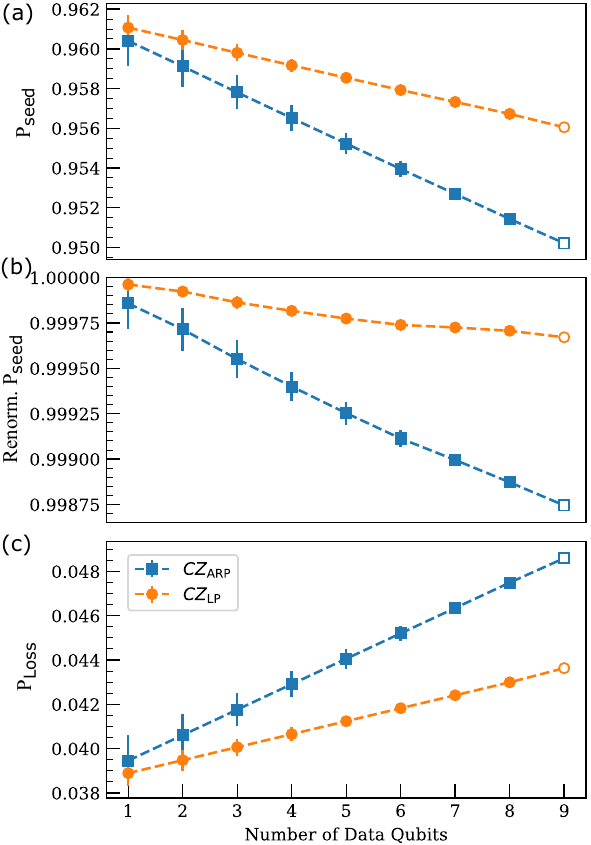}
    \caption{\label{fig:BV}{The results for the Bernstein-Vazirani algorithm simulations are given for both $CZ$ gate implementations run from $1+1$ to $9+1$ qubits. Unfilled data points denote cases where the circuit sample size was limited to sample only 256 rather than all possible circuits. (a) The raw success probability $P_{\mathrm{Seed}}$ and (b) the post-selected success probability. (c) The loss probability $P_\mathrm{Loss}$.}{\color{black}}}
\end{figure}

As a means of testing the effect of the dynamic connectivity procedure on the performance of a specific algorithm we implement the Bernstein-Vazirani (BV) algorithm \cite{bernstein97}. This algorithm uses an oracle to identify an $N$-bit binary string using $N$ measured (data) qubits with one {\color{black}un}measured (ancilla) qubit {\color{black}, and requiring a single measurement, thus} representing a polynomial speedup compared to the $N$ operations necessary to perform this task using a classical algorithm \cite{londero04}.

Multiple \km{realisations} of this algorithm are available \cite{arvind07} but the specific implementation {\color{black}that} we consider makes use of an oracle which exploits entanglement between the data and ancilla qubits through CNOT gates applied to the ancilla qubit conditioned on each data qubit where the index corresponds to a 1 in the bit string. \km{The oracle enacts a function which takes the dot product of the system state $x$ and an encoded binary secret seed $s$, of the form:
\begin{equation}
    |x\rangle\rightarrow(-1)^{(x \cdot s)}|x\rangle.
\end{equation}
The measurement output of the data qubits in the computational basis should then match the $N$-bit binary seed string $s$, meaning the algorithm has correctly identified the string.}

This formulation of the BV algorithm lends itself well to the use of qubit movement due to the simple star-shaped connectivity requirement, where the only necessary connections are between each data qubit and the ancilla qubit. 

We analyse at a 9+1 qubit device to assess performance of this problem in the regime of reconfigurable connectivity, making use of qubit zoning to minimize the number of moves necessary, as demonstrated in Fig.~\ref{fig:BVInfo}(b).

As the number of possible bit strings increases exponentially with qubit number, we cap the number of bit strings considered at 256 to facilitate simulation. This provides an approximation of the results at higher circuit sizes. In order to mitigate any bias toward these circuit sizes we consider a random sample of 254 bit strings, and always consider the smallest seed string $00\cdots0$ and the largest seed string $11\cdots1$.

The results of the simulations for between 1+1 and 9+1 qubit implementations of the algorithm are given in Fig.~\ref{fig:BV} and demonstrate that this algorithm can be efficiently implemented on Rydberg atom devices with a high success rate for 9+1 qubits. In this case the LP protocol gives raw probabilities over 0.95 and relatively low loss probabilities ($< 0.05$) due to the significant reduction in two-qubit gate count compared to the quantum volume metric above. With renormalization $>0.999$ success probability can be achieved, where here the ARP gates still perform worse than LP highlighting the sensitivity of different algorithms to different dominant gate errors and showing that the BV circuits are more tolerant to small phase-errors in the two-qubit gates. These results highlight the capability for scaling to larger qubit numbers for implementing the BV algorithm. These results also demonstrate a marked improvement compared to previous work experimentally benchmarking this algorithm on small-scale superconducting and trapped-ion devices by Linke \emph{et al.} \cite{linke17}\km{, reflecting the significant increase in coherence times and gate fidelities in the time since this work was completed}. 

\subsection{Grover's Search Algorithm} 

\begin{figure}[t!]
     \centering    
         \includegraphics[width=\columnwidth,keepaspectratio]{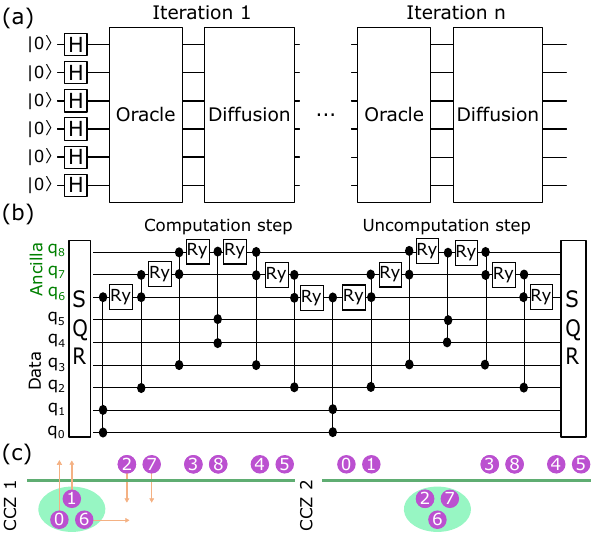}
    \caption{\label{fig:GroversInfo}{Grover's search algorithm. (a) The form of the circuit we consider to implement Grover's search algorithm. This consists of n iterations, each consisting of an application of the oracle and diffusion operator. (b) An example of the concatenated $CCZ$ decomposition of the $C_{5}Z$ gate required for the construction of the oracle and diffusion operators, where these are {\color{black}book-ended by} the necessary single qubit rotations (SQR). This procedure uses 3 ancilla qubits to realise the $C_{5}Z$ gate. (c) Proposed movement procedure for implementing Grover's algorithm using $CCZ$ gates.}}
\end{figure}

\km{Grover's search algorithm is a primary example of a useful application of quantum computing. This algorithm is a fast way to perform an unstructured search, offering a $\sqrt{N_{s}}$ speedup for an $N_{s}$ item search space compared to classical algorithms \cite{nielsen11}, and makes use of entanglement between all qubits to do so. This requirement of multi-qubit entanglement makes this algorithm a natural choice to test the effect of native multi-qubit gates on the algorithmic performance on Rydberg atom systems.}

\km{We implement Grover's algorithm through repeated applications of an oracle, which takes the form of an identifying phase on the target state, and diffusion operator, which takes the form of a phase flip on each basis state except $\ket{0\cdots0}$. These two operators combined constitute a single iteration step, which result in a rotation of the state of the data qubit register Hilbert space toward the oracle-encoded target state \cite{nielsen11}. Each successive iteration brings the system state closer to the target state hence amplifying the output state measurement probability.} 

We consider a circuit construction making use of an identifying phase as the oracle \cite{nielsen11}, as shown in Fig.~\ref{fig:GroversInfo}(a). This has an intuitive implementation on Rydberg atom devices \cite{molmer11}, as the native $CCZ$ gates allow for an implementation of both this oracle and the diffusion operator through a procedure based on concatenated Toffoli gates with ancilla qubits \cite{barenco95} which we translate into the native $CCZ$ gate, as illustrated in Fig.~\ref{fig:GroversInfo}(b) to provide a 6-qubit search algorithm on a 9-qubit device. 

\begin{figure}[t!]
     \centering    
         \includegraphics[width=\columnwidth,keepaspectratio]{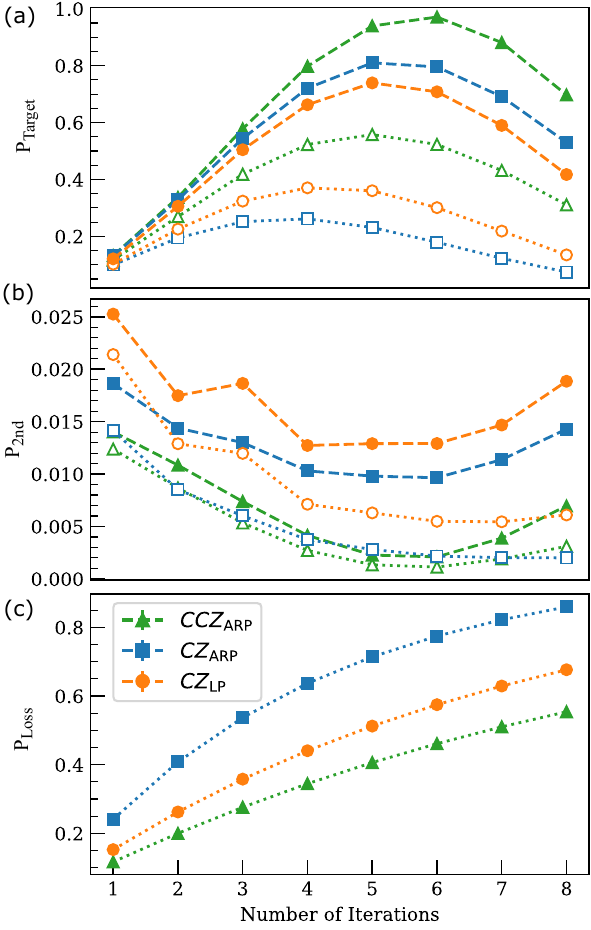}
    \caption{\label{fig:Grovers}{The results for the Grover's algorithm simulations are presented for all three implementation schemes in a reconfigurable processor. (a) The raw (unfilled) and post-selected (filled) success probabilities after each iteration step. (b) The probabilities for the next-most-probable output. (c) The loss probability $P_{\mathrm{Loss}}$}.}
\end{figure}

For this 6+3 qubit circuit, we define a movement protocol illustrated in Fig~\ref{fig:GroversInfo}(c) to perform successive $CCZ$ gate operations between groups of qubits, \km{where in each stage a trio of qubits are moved to provide the necessary connectivity to implement the next required $CCZ$ gate.} Additionally, we compare against two-qubit gate decomposition's using both ARP and LP gate protocols. \km{The circuits we consider are manually transpiled to attempt to minimise the number of gates used in the implementation, however an efficient automated transpilation routine tailored to the system specifications would likely reduce the number of gates and hence minimise the error incurred in the circuit.} 

To benchmark performance we simulate the algorithm for all possible outputs in the search space for increasing numbers of oracle queries to see how errors in imperfect gates negate improvements in output state amplification from successive oracle applications. The results of these simulations are shown in Fig.~\ref{fig:Grovers}, where we {\color{black}show} the probability of observing the target output state, $P_\mathrm{Target}$, and the probability associated with the next most likely output, $P_\mathrm{2nd}$, to {\color{black}facilitate analysis of the ability to identify the target state from the distribution}.

Here the $CCZ$ implementation yields raw output probabilities exceeding 0.50 after 5 iterations, whilst achieving the lowest loss probability and largest suppression of the incorrect output state amplitudes below 0.002. By comparison, for the LP and ARP $CZ$ gate circuits the raw output probabilities peak after only 4 iterations at around 0.35 and 0.25 respectively, {\color{black}with} reduced suppression of the next-most likely output and, in the case of the ARP $CZ$ gate, significant losses.  

In the regime of loss-correction, the output probabilities peak an iteration higher, with the corrected $CCZ$ circuit yielding 0.97 success probability, whilst in the $CZ$ based circuits the ARP protocol outperforms LP due to reduced phase errors but with a peak success probability closer to only 0.80. These results clearly demonstrate the advantages of being able to apply native multi-qubit gates for efficient and high-fidelity implementation of quantum algorithms.

\section{Discussion and Outlook}
\label{sec:discussion}
In this paper we have analyzed the theoretical performance of near-term neutral atom quantum processors {\color{black}by} using algorithmic benchmarks and estimating the attainable quantum volume, comparing different operation modalities and gate protocols. 

Comparison of quantum volume for circuits up to a depth of 9 layers on 9 qubits shows that significantly higher performance can be obtained using dynamically reconfigurable qubit connectivity instead of relying on static qubit geometries. Further, we show that it is important to consider not only the absolute gate fidelity, but also the dominant error channels, where the low loss of the LP protocol yields improved output probabilities compared to the low-phase error ARP $CZ$ gates {\color{black}unless post-processing is possible, in which case this comparison is in some instances reversed}. For moveable tweezers using the LP gate protocol we estimate  a lower bound of $V_Q=2^9$ without loss correction. \km{With post-selection we see both gate schemes reach $V_Q=2^9$, with a higher heavy output probability for the ARP $CZ$ gate. In both cases movement based connectivity represents an improvement on static array SWAP gate protocols.}

{\color{black}By simulation of paradigmatic algorithm implementations on Rydberg quantum computers} we have shown {\color{black}that} the 9+1 qubit BV algorithm can yield raw success probabilities $>0.95$, {\color{black}and that} exploiting the native implementation of multi-qubit $CCZ$ gates provides an efficient approach to Grover's search algorithm, significantly increasing performance and suppressing losses compared to a $CZ$ gate based decomposition{\color{black}, which many other hardwares are limited to}.

Through this work we have demonstrated that current neutral atom quantum computers are capable of efficient, high quality algorithmic implementation in the NISQ era, with the predicted quantum volume metrics highly competitive against current superconducting circuit systems. Future work to extend these simulations to larger qubit numbers, combined with implementation of improved circuit transpilation and efficient routing of atom movements \cite{tan23} will enable accurate estimation of the limiting quantum volume for larger systems, along with demonstration of experimental verification using real hardware.

\emph{Note:} Whilst preparing this manuscript we became aware of similar work by Wagner \emph{et al.} \cite{wagner23} performing numerical simulations of algorithmic performance of up to 11 qubits based on the experimental hardware from Ref.~\cite{graham22}, demonstrating a 10-15\% improvement \km{in the average success probability of a number of algorithms} in the limit of all-to-all connectivity. 

\begin{table}[t!]
    \centering
    \begin{tabular}{|c|c|}
    \hline
    Simulation Parameters & Value \\
    \hline
         $T_{1}$ \cite{bluvstein22}& 4.00 s \\
    \hline
         $T_{2}$ \cite{bluvstein22}&  1.49 s \\
    \hline
      $\Omega$ \cite{bluvstein22}& 1 MHz\\
    \hline
    $\epsilon_\mathrm{Init}   ~\cite{evered23}$&0.003  \\
     \hline
    $t_\mathrm{Init}$ \cite{levine19} & $300~\mu$s\\
    \hline
   $v_\mathrm{move}$ \cite{bluvstein22} & $0.55~\mu $m$\,\mu$s$^{-1}$ \\
    \hline
     $\tau_\mathrm{move}^\mathrm{A}$  & $100~\mu$s \\
     \hline
     $\tau_\mathrm{move}^\mathrm{B}$  & $40~\mu $s \\
    \hline
    \end{tabular}
    \caption{Default simulation parameters for the Rydberg Virtual Quantum Device simulated in QuEST. Other input parameters are bespoke or irrelevant to the tasks considered in this work }
    \label{tab:param}
\end{table}

\section{Acknowledgements}
The authors would like to acknowledge the QuEST team, particularly C. Gustiani, for development of and providing early access to the virtual quantum device package on which these simulations were performed. We also thank Andrew Daley, Viv Kendon and Gerard Peligri for useful discussions and careful reading of the manuscript. This work is supported by the EPSRC Prosperity Partnership SQuAre (Grant No. EP/T005386/1) with funding from M Squared Lasers Ltd. The data presented in this work are available at \cite{mckinroy23data}.


\begin{thebibliography}{10}
\bibitem{mckinroy23data}
{DOI to be added in proof}.
\end{thebibliography}

\end{document}